\documentclass[twocolumn,showpacs,preprintnumbers,amsmath,amssymb]{revtex4}

\usepackage{graphicx}
\usepackage{dcolumn}
\usepackage{bm}

\begin{document}

\preprint{Eur. Phys. J. B 59, 249(2007)}

\title{Frequency and phase synchronization of two coupled neurons with channel noise}

\author{Lianchun Yu, Yong Chen\footnote[1]{Corresponding author. Email: {\tt ychen@lzu.edu.cn}}, and Pan Zhang}

\address {Institute of Theoretical Physics, Lanzhou University, Lanzhou $730000$, China}

\date{\today}

\begin{abstract}
We study the frequency and phase synchronization in two coupled
identical and nonidentical neurons with channel noise. The
occupation number method is used to model the neurons in the context
of stochastic Hodgkin-Huxley model in which the strength of channel
noise is represented by ion channel cluster size of neurons. It is
shown that channel noise allows the two neurons to achieve both
frequency and phase synchronization in the regime where the
deterministic Hodgkin-Huxley neuron is unable to be excited. In
particular, the identical channel noises lead to frequency
synchronization in weak-coupling regime. However, if the coupling is
strong, the two neurons could be frequency locked even though the
channel noises are not identical. We also show that the relative
phase of neurons displays profuse dynamical regimes under the
combined action of coupling and channel noise. Those regimes are
characterized by the distribution of the cyclic relative phase
corresponding to antiphase locking, random switching between two or
more states. Both qualitative and quantitative descriptions are
applied to describe the transitions to perfect phase locking from no
synchronization states.
\end{abstract}


\maketitle

\section{\label{sec1}INTRODUCTION}

The synchronization phenomena have been widely studied in neural
systems in past decades \cite{kurth,Gray2,Robert,Pinto,li02}.
Experiments show that the synchronization of coupled neurons could
play a key role in the biological information communication of
neural systems \cite{Gray2}. Recent research also suggests that
synchronization behavior is of great importance for signal encoding
of ensembles of neurons. Especially the phase synchronization may be
important in revealing communication pathways in brain
\cite{Fitzgerald}. Studying the synchronization of a pair of coupled
neurons has attracted large amounts of research attention. In order
to understand the dynamical properties of a neural network, it is
important to characterize the relation between spike trains of two
neurons in the network \cite{Uchida}. What's more, studies show that
noise enhances synchronization of neural oscillators. For example,
the identical neurons which are not coupled or weakly coupled but
subjected to a common noise may achieve complete synchronization.
Actually, this is a general results for all the dynamical system
\cite{pikovsky1,Maritan,pikovsky2, zhou,Herzel}. Both independent
and correlated noises are found to enhance phase synchronization of
two coupled chaotic oscillators below the synchronization threshold
\cite{Yuqing Wang, Istvan}.

Among large population of neurons, different neurons are commonly
connected to other group of neurons and receive signals from them.
As a result of integration of many independent synaptic currents,
those neurons receive a common input signal which often approaches a
Gaussian distribution \cite{Changsong Zhou}. Therefore, noise was
usually considered as external and introduced by adding to the input
variables. However, recent work found that the random ionic-current
changes produced by probabilistic gating of ion channels, called
channel noise or internal noise, also play an amazing role in single
neuron's firing behavior and information processing progress
\cite{Hille,White1,Freedman}. Besides, Casado has showed that
channel noise can allow the neurons to achieve both frequency and
phase synchronization \cite{Casado1, Casado2}. This finding suggests
that channel noise could play a role as promoter of synchronous
neural activity in population of weakly coupled neurons. However,
Casado didn't give a quantitative description of the results.

The magnitude of the ion channel noise is changed via the variation
of the channel cluster size of neurons. It implies that
synchronization in neural system is also restricted by the channel
cluster size of neurons. Actually, the cluster size of ion channels
embedded in the biomembrane between the hillock and the first
segment of neurons determines whether the neuron fires an action
potential. The channel cluster size of this region is different for
different neurons or for different developing stages of a neuron.
With the decreasing of ion channel cluster size, ion channel noise
would be induced thus the firing behavior of neurons would be
greatly changed (for review see Ref\cite{White1}). It is natural to
ask to what extent the change of the channel cluster size affect the
collective activities of neuron ensembles. For this purpose, we
investigated the effect of ion channel cluster sizes (i.e, channel
noise ) of neurons on synchronization of two coupled stochastic
Hodgkin-Huxley (HH) neurons in this paper.

Here we adopted a so called occupation number method rather than the
Langevin method Casado had used to describe the single neuron for
two reasons. First, the Langevin approach has been proved could not
reproduce accurate results for small and large cluster sizes.
Second, occupation number method gives a direct relation between
channel cluster size of neuron and its firing behavior, and it's the
fastest method for a given accuracy \cite{Shangyou Zeng}. The main
goal of our work is to explore what role the channel noise might
play in the synchronization of two coupled neurons. We try to give
qualitative as well as quantitative descriptions of the result. The
practical meaning of our study is obvious, since the channel cluster
size of neurons can be regulated by channel blocking experimentally
\cite{Schmid}, our study may provide a possible way to control
neural synchronization.

This paper is organized as follows. In section \ref{sec2}, the
occupation number method of stochastic Hodgkin-Huxley neuron is
introduced and the firing behaviors of neurons with different
channel cluster sizes are demonstrated. In the following sections,
we explored the combined effect of coupling strength and cluster
size on the synchronization behaviors of two neurons with an
electrical synaptic connection. Section \ref{sec3} is devoted to
frequency synchronization. The phase synchronization of identical
and nonidentical neurons are discussed in Section \ref{sec4}. A
conclusion is presented in Section \ref{sec5}.

\section{The model\label{sec2}}

Hodgkin-Huxley neuron model provides direct relation between the
microscopic properties of ion channel and the macroscopic behavior
of nerve membrane. The membrane dynamics of HH equations is given by
\begin{eqnarray}
C_{m}\frac{dV}{dt} &=& -(G_{K}(V-V_{K}^{rev})+ \;G_{Na}(V-V_{Na}^{rev})\nonumber\\
 & &+\;G_{L}(V-V_{L})-\,I),
\label{eq-1}
\end{eqnarray}
where $V$ is the membrane potential. $V_{K}^{rev}$, $V_{Na}^{rev}$,
and $ V_{L}$ are the reversal potentials of the  potassium, sodium,
and leakage currents, respectively. $G_{K}$, $G_{Na}$, and $G_{L}$
are the corresponding specific ion conductances. $C_{m}$ is the
specific membrane capacitance, and $I$ is the current injected into
this membrane patch. The voltage-dependent conductances for the
$K^{+}$ and $Na^{+}$ channels are given by
\begin{equation}
G_{K}=\gamma_{K}\frac{N^{open}_{K}}{S} \;,\;
       G_{Na}=\gamma_{Na}\frac{N^{open}_{Na}}{S}\;,
 \label{eq-2}
\end{equation}
where $N^{open}_{K}$ and $N^{open}_{Na}$ are the numbers of open
potassium and sodium channels. $S$ is the membrane patch area.
$\gamma_{K}$ and $\gamma_{Na}$ give the single-channel conductances
of $K^{+}$ and $Na^{+}$ channels. Then the numbers of total
potassium and sodium  channels $N_{K}$ and $N_{Na}$ are given by the
equations $N_{K}=\rho_{K}\times S $ and $N_{Na}=\rho_{Na}\times S $,
where $\rho_{K}$ and $\rho_{Na}$ are the $K^{+}$ and $Na^{+}$
channel densities respectively. By introducing time constants
$\tau_{K}=\frac{C_{m}}{\rho_{K}\gamma_{K}}$,
$\tau_{Na}=\frac{C_{m}}{\rho_{Na}\gamma_{Na}}$ and
$\tau_{L}=\frac{C_{m}}{G_{L}}$, we end up with the following
equation for the membrane potential
\begin{eqnarray}
\frac{dV}{dt} &=&-(\frac{N_{K}^{open}}{\tau_{K}N_{K}}(V-V_{K}^{rev}) + \; \frac{N_{Na}^{open}}{\tau_{Na}N_{Na}}(V-V_{Na}^{rev})\nonumber\\
& &+\;\frac{1}{\tau_{L}}(V-V_{L})-I)\;.
\label{eq-3}
\end{eqnarray}

Individual channels open and close randomly. If the number of
channels are large and they act independently of each other, then,
from the law of large numbers, $N_{K}^{open}/\;N_{K}$ ( or
$N_{Na}^{open}/N_{Na}$) is approximately equal to the probability
that any one $K^{+}$ (or $Na^{+}$) channel is in an open state, and
can be represented as continuous deterministic gating variables
$n^{4}$ and $m^{3}h$. This leads to the deterministic version of HH
model \cite{Hodgkin-Huxley,Abbott},
\begin{eqnarray}
C_{m}\frac{dV}{dt} &=& -\overline{g}_{K}n^{4}(V-V_{K}^{rev}) -\; \overline{g}_{Na}m^3h(V-V_{Na}^{rev})\nonumber\\
& &-\;G_{L}(V-V_{L})+ I,
 \label{eq-4}
\end{eqnarray}
where $\overline{g}_{K}=\rho_{K}\times\gamma_{K}$ and
 $\overline{g}_{Na}=\rho_{Na}\times\gamma_{Na}$ are the maximal potassium
and sodium conductance per unit area. $n^{4}$ indicates that the
$K^{+}$ channel has four separate gates and that a $K^{+}$ channel is
opened when all those gates are open; $m^{3}h$ indicates that three
$m$-gates and one $h$-gate must be opened to open a $Na^{+}$
channel. The gating variables obey the following equations,
\begin{equation}
\frac{d}{dt}x=\alpha_{x}(V)(1-x)-\beta_{x}(V)x, x=m, h, n,
\label{eq-5}
\end{equation}
where $\alpha_{x}(V)$ and $\beta_{x}(V)$ ($x=m, h, n$) are voltage
dependent opening and closing rates and are given in Table
\ref{table-1} with other parameters used in the simulations.

\begin{table} [t]
\caption{Parameters and Rate Functions Used in Simulations.}
\begin{tabular*}{8.5cm}{@{\extracolsep{\fill}}lll}\\
\hline
                &                                  &                       \\
\small$ C_{m}       $  &\small Specific membrane capacitance    &\small $1\mu F / cm^{2}$ \\
\small$V^{rev}_{K}  $  &\small Potassium reversal potential     &\small $ -77mV        $ \\
\small$ V^{rev}_{Na}$  &\small Sodium reversal potential        &\small $50mV$ \\
\small$ V_{L}       $  &\small Leakage reversal potential       &\small $-54.4mV$ \\
\small$ \gamma_{K}  $  &\small Potassium channel conductance    &\small $20pS$\\
\small$ \gamma_{Na} $  &\small Sodium channel conductance       &\small $20pS$\\
\small$ G_{L}       $  &\small Leakage conductance              &\small $0.3mS/cm^{2}$\\
\small$ \rho_{K}    $  &\small Potassium channel density        &\small $20/\mu m^{2}$\\
\small$ \rho_{Na}   $  &\small Sodium channel density           &\small $60/\mu m^{2}$\\
\small$ \tau_{K}    $  &\small Potassium channel time constant  &\small $1/36 ms   $\\
\small$ \tau_{Na}   $  &\small Sodium channel time constant     &\small $1/120 ms   $\\
\small$ \tau_{L}    $  &\small Leakage channel time constant    &\small $3.3 ms   $\\
\small$ \alpha_{n}  $  &                          &$\frac{0.01(V+55)}{1-e^{-(V+55)/10}}$\\
\small$\beta_{n}    $  &                          & $0.125e^{-(V+65)/80} $\\
\small$\alpha_{m}   $  &                          & $\frac{0.1(V+40)}{1-e^{-(V+40)/10}} $\\
\small$\beta_{m}    $  &                          & $4e^{-(V+65)/18}$\\
\small$\alpha_{h}   $  &                          & $ 0.07e^{-(V+65)/20}$\\
\small$\beta_{h}    $  &                          & $ \frac{1}{1+e^{-(V+35)/10}}$\\
                       &                          &                           \\
\hline
\end{tabular*}
\label{table-1}
\end{table}
\begin{figure}
\includegraphics[width=0.50\textwidth]{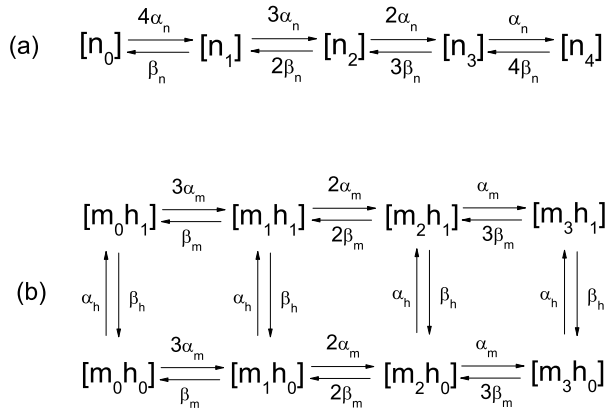}
\caption{Kinetic scheme for a stochastic potassium channel (a) and
sodium channel (b). $n_{4}$ and $m_{3}h_{1}$ are open states, while
the other states are no-conducting. } \label{fig:1}
\end{figure}

The deterministic HH neuron model [Eqs. (\ref{eq-4})-(\ref{eq-5})]
describes the transmembrane potential without the need to treat the
underlie activity of individual ion channels. However, for the
limited number of channels, Eq. (\ref{eq-4}) is no longer valid and
statistical fluctuations will play a role in neuronal
dynamics\cite{White1}. So we have to return to Eq. (\ref{eq-3}) and
have to determine $N^{open}_{K}$ and $N^{open}_{Na}$ as a function
of time by stochastic simulation methods based on state diagrams
that indicate the possible conformation states of channel molecules.

As shown in Fig.~\ref{fig:1}, both $K^{+}$ and $Na^{+}$ channels
exist in many different states and switch between them according to
voltage depended transition rates (identical to the original HH rate
functions). $[n_{i}]$ is the state of $K^{+}$ channel with $i$ open
gates and $[m_{i}h_{j}]$ is the state of $Na^{+}$ channel with $i$
open $m$-gates and $j$ open $h$-gates. Hence, $[n_{4}]$ labels the
single open state of the $K^{+}$ channel and $[m_{3}h_{1}]$ the
$Na^{+}$ channel. Usually we can simulate the kinetic scheme of each
ion channel to get the numbers of open sodium and potassium channels
at each instant. However, it is not an efficient way because many
transitions of states do not change the conductance of the channel.
Instead of keeping track of the state of each channel, we keep track
of the total populations of channels in each possible state so
$N_{K}^{open}$ and  $N_{Na}^{open}$ at each instant can simply be
determined by counting the numbers of channels in state $[n_{4}]$
and $[m_{3}h_{1}]$. Specifically, if the transition rate between
state $A$ and state $B$ be $r$ and the number of channels in these
states be $n_{A}$ and $n_{B}$. Then, the probability that a channel
switches within the time interval ($t, t+\Delta t$) from state $A$
to $B$ is given by $p=r\Delta t$. Hence, for each time step, we
determine $\Delta n_{AB}$, the number of channels switch from $A$ to
$B$, by choosing a random number from a binomial distribution
\cite{Freedman},
\begin{equation}
P(\Delta n_{AB})=\left(^{n_{A}}_{\Delta n_{AB}}\right)p^{\Delta
n_{AB}} (1-p)^{(n_{A}- \Delta n_{AB})}.
\end{equation}
Then we update $n_{A}$ with $n_{A}-\Delta n_{AB}$, and $n_{B}$ with
$n_{B}+\Delta n_{AB}$. To make sure that the number of channels in
each state is positive, we update those number sequentially,
starting with the process with the largest rate and so forth.
\begin{figure}
\includegraphics[width=0.50\textwidth]{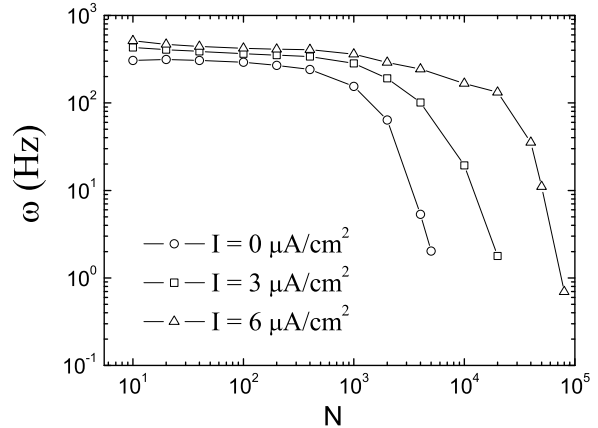}
\caption{The mean firing frequency as a function of the channel
cluster size $N$ for $I=0 ~\mu A/cm^{2}(\circ)$, $I=3 ~\mu
A/cm^{2}(\square)$ and $I=6 ~\mu A/cm^{2}(\vartriangle)$. The data
are obtained from spike trains of 2000 action potentials. }
\label{fig:2}
\end{figure}

   Voltage-gated ion channels are stochastic devices. The origin of channel noise is
basically due to fluctuations of the fraction of open ion channels
(thus the channel currents) around the corresponding mean values.
The strength of the fluctuation is inversely proportional to the
number of total ion channels\cite{White1, Chow}. Though average
membrane current is at constant, as we will see, the variance of the
$Na^{+}$ and $K^{+}$ currents cause remarkable effects on neuronal
dynamics. In this work, we introduce channel cluster size $N$
($N=N_{K}=N_{Na}/3$) as a measurement of channel noise level so that
the correct proportion between $Na^{+}$ and $K^{+}$ channel
densities is preserved. With increasing channel cluster size, the
fluctuations of the fraction of open ion channels, thus the variance
of the the corresponding channel currents decreases. For a large
number of channels this noise becomes negligible (i.e., the
deterministic case). The threshold constant current for
deterministic HH neuron to generate consecutive action potentials is
$I_{th}=6.26 \mu A/ cm^{2}$. However, due to the channel noise, the
stochastic HH neurons can generate spiking activity with
subthreshold input current\cite{Shangyou Zeng}. Fig.~\ref{fig:2}
shows the mean firing frequency (defined in Sec. \ref{sec3}) as a
function of channel cluster size for different constant current. If
the channel cluster size is small, the neuron fires action
potentials with high frequency. As the channel cluster size
increases, the mean firing frequency drops quickly, approaching the
deterministic case that no firing activities occur with the same
subthreshold input currents. With decreasing the input current, the
firing frequency decreases. However, the firing activities will not
vanish if the input current is decreased to zero. Thus, as have
demonstrated, channel noise shifts the onset of firing behavior to
lower values of input current $I$.

 To explore the synchronization phenomena, we
consider two stochastic HH neurons coupled by an electrical synaptic
connection. The system is described by the following equations,
\begin{widetext}
\begin{eqnarray}
\frac{dV_{1}}{dt}&=&-\frac{N_{K_{1}}^{open}}{\tau_{K}N_{K_{1}}}(V_{1}-V_{K}^{rev}) - \frac{N_{Na_{1}}^{open}}{\tau_{Na}N_{Na_{1}}}(V_{1}-V_{Na}^{rev}) - \frac{1}{\tau_{L}}(V_{1}-V_{L})   +\varepsilon(V_{1}-V_{2})+I_{1}, \label{eq-6}\\
\frac{dV_{2}}{dt}&=& -\frac{N_{K_{2}}^{open}}{\tau_{K}N_{K_{2}}}(V_{2}-V_{K}^{rev}) -\frac{N_{Na_{2}}^{open}}{\tau_{Na}N_{Na_{2}}}(V_{2}-V_{Na}^{rev}) -\frac{1}{\tau_{L}}(V_{2}-V_{L})+\varepsilon(V_{2}-V_{1})+I_{2}.\label{eq-7}
\end{eqnarray}
\end{widetext}
Here $V_{1}$ and $V_{2}$ are the instantaneous membrane potentials
of the two neurons and $\varepsilon$ is the diffusive coupling
strength between the neurons. $N_{K_{1}}^{open}$,
$N_{Na_{1}}^{open}$, $N_{K_{2}}^{open}$, $N_{Na_{2}}^{open}$ are the
numbers of open $K^{+}$ and $Na^{+}$ channels of neuron $1$ and $2$,
respectively ; $N_{K_{1}}$, $N_{Na_{1}}$, $N_{K_{2}}$, $N_{Na_{2}}$
are the numbers of total $K^{+}$ and $Na^{+}$ channels for neuron
$1$ and $2$, respectively. $I_{1}$ and $I_{2}$ are two constant
input currents which are set at $I_{1}=I_{2}=6\mu A/cm^{2}$. Here,
$N_{1}$ and $N_{2}$ ($N_{i}=N_{K_{i}}=N_{Na_{i}}/3,\; i=1, 2$) are
the channel cluster sizes for each neuron.

The numerical integration of system mentioned above is carried out
by using the Euler algorithm with a step size of $0.01$ms. And all
simulations are working in Ito framework. The occurrences of action
potentials are determined by upward crossings of the membrane
potential at a certain detection threshold of 10mV if it has
previously crossed the reset value of -50mV from below.

\section{frequency synchronization\label{sec3}}

\begin{figure}
\includegraphics[width=0.5\textwidth]{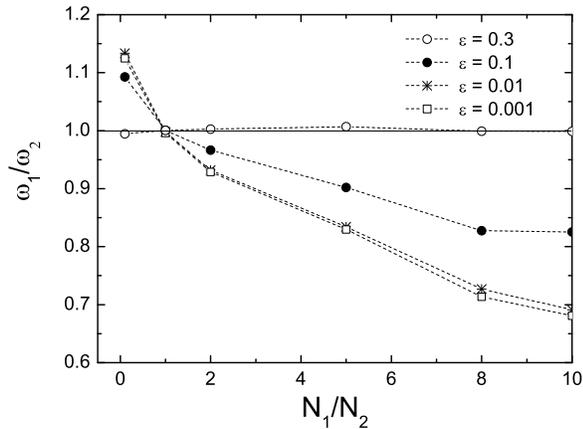}
\caption{The winding number $\omega_{1}/\omega_{2}$ as a function of
$N_{1}/N_{2}$ with $N_{2}=2\times 10^{2}$ and $\varepsilon=0.3$ ($\circ$),
$\varepsilon=0.1$ ($\bullet$), $\varepsilon=0.01$ ($\ast$),
$\varepsilon=0.001$ ($\square$). }
\label{fig:3}
\end{figure}

\begin{figure*}[t]
\includegraphics[width=0.85\textwidth]{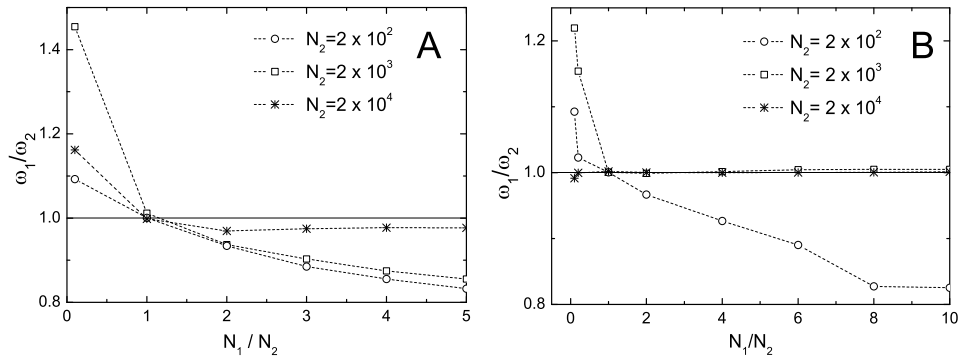}
\caption{The winding number $\omega_{1}/\omega_{2}$ as a function of
$N_{1}/N_{2}$ with $\varepsilon=0.01$ (A), $\varepsilon=0.1$ (B) and
different $N_{2}$: $N_{2}=2\times10^{2}$ ($\circ$),
$N_{2}=2\times10^{3}$ ($\square$) and $N_{2}=2\times10^{4}$
($\ast$).} \label{fig:4}
\end{figure*}

From the above-mentioned system, we can get two point processes
in the following form,
\begin{equation}
z(t)=\sum_{n=1}^{N}\delta(t-t_{n}).
\label{eq-8}
\end{equation}
Each one gives the spike sequence of a particular neuron. The mean
spiking frequency of neuron $i$ ($i=1, 2$) is defined as,
\begin{equation}
\omega_{i}=\lim_{N \ \rightarrow
\infty}\frac{1}{N}\sum^{N}_{n=1}\frac{2\pi}{t_{n+1}-t_{n}}, \;  i=1, 2.
\label{eq-9}
\end{equation}
Generally, synchronization means an adjustment of timescales of
oscillations in systems due to their circumstances. In other words,
oscillators can shift the timescales to make their ratio close to a
rational number $n:m$, where $n$ and $m$ are integers. This
phenomenon is usually referred to as $n:m$ frequency
synchronization, and its suitable measure is the closeness of the
ratio of $\omega_{1}/\omega_{2}$ to the chosen rational number $n:m$
\cite{Hauschild}. In this paper we will discuss only $1:1$
synchronization. Note that the frequency locking discussed in this
section is in a stochastic sense, and refers to the equivalence of
the average frequencies rather than the instantaneous frequency. So
it is not a sufficient condition for synchronization. However, since
the firing rate of a neuron is often argued to carry information of
the stimulus, studying of the frequency synchronization is
especially meaningful in the context of rate coding scheme.

We investigated the shift of winding number $\omega_{1}/\omega_{2}$
along with both the variation of coupling strength and channel
cluster size. When the coupling strength is small, as shown in
Fig.~\ref{fig:3}, with the increasing of $N_{1}/N_{2}$ from
$N_{1}/N_{2}<1$, the winding number will decrease from a value at
which the two neurons are not synchronized. When two channel cluster
sizes are the same ( $N_{1}/N_{2}=1$), both neuron will be frequency
locked. Further increasing of $N_{1}/N_{2}$ will desynchronize them
to a certain levels. Note that in the range of $N_{1}/N_{2}>1$,
increasing the value of coupling strength tends to increase
$\omega_{1}/\omega_{2}$, and the increasing is larger with larger
coupling strength. In panel A of Fig.~\ref{fig:4}, it is seen
clearly that though frequency synchronization can be achieved with
arbitrary chosen value of $N_{2}$, the tuning is very critical, as a
small variation of $N_{1}$ from $N_{2}$ leads to desynchronization.
As has been described above, for a isolated neuron, its average
firing rate decreases with increasing its channel cluster size
(i.e., decreasing the channel noise intensity). Thus, at a given
channel cluster size, the channel noise term is identical for the
two neurons. In this case, their average firing rates would be the
same, which means the two neurons are frequency locked.

However, if the coupling strength is increased to a rather large
value (for example, $\varepsilon=0.3$ in Fig.~\ref{fig:3}), the
coupling strength starts to take command of the frequency
synchronization as the two neurons are able to be entrained in a
wider range of channel cluster size (i.e., channel noise level). It
is seen in panel B of Fig.~\ref{fig:4} that neurons with larger
value of $N_{2}$ is easier to get frequency entrained in a wider
range with lower coupling strength. Even in the weak-coupling case,
as shown in panel A of Fig.~\ref{fig:4}, large channel cluster sizes
tend to enhance synchronization (make $\omega_{1}/\omega_{2}$ closer
to 1). This implies that neurons with large channel cluster sizes
(i.e., small channel noise level) are easier to adjust their
timescales to make their firing rates close to each other. However,
if the channel noises are too small, the neurons won't fire spikes
with subthreshold stimuli.

It is concluded  that for identical, symmetrically coupled neurons,
when the coupling strength is small, the channel cluster sizes at
frequency synchronization must be the same, whereas the coupling
strength only has a limited effect only when the channel cluster
sizes of the two neurons are not same. However, when the coupling
strength is rather large, the two neurons are able to achieve
frequency synchronization with a greater range of channel cluster
sizes. In this regime, though large channel noise degrade frequency
synchronization, small channel noise intensities help to get the
neurons frequency synchronized with subthreshold stimuli.

\section{phase synchronization\label{sec4}}

\begin{figure}
\includegraphics[width=0.5\textwidth]{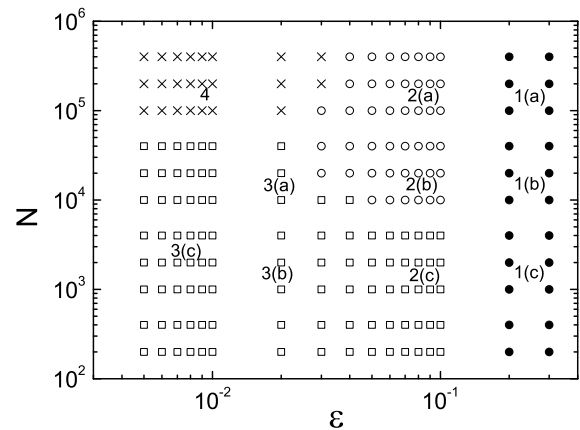}
\caption{Synchronization diagram for the distribution of cyclic relative phase $P(\Phi)$. Region 1 correspond to state of monomodal distribution ($\bullet$), region 2 to bimodal distribution ($\circ$), region 3 to drifting evolution of $\Delta\phi$ ($\square$); region 4 to no firing area ($\times$). There are no lines plotted to separate the subregions in those regions because the distribution of $P(\Phi)$ changes in a continuous manner in those regions. }
\label{fig:5}
\end{figure}

\begin{figure*}[t]
\includegraphics[width=0.8\textwidth]{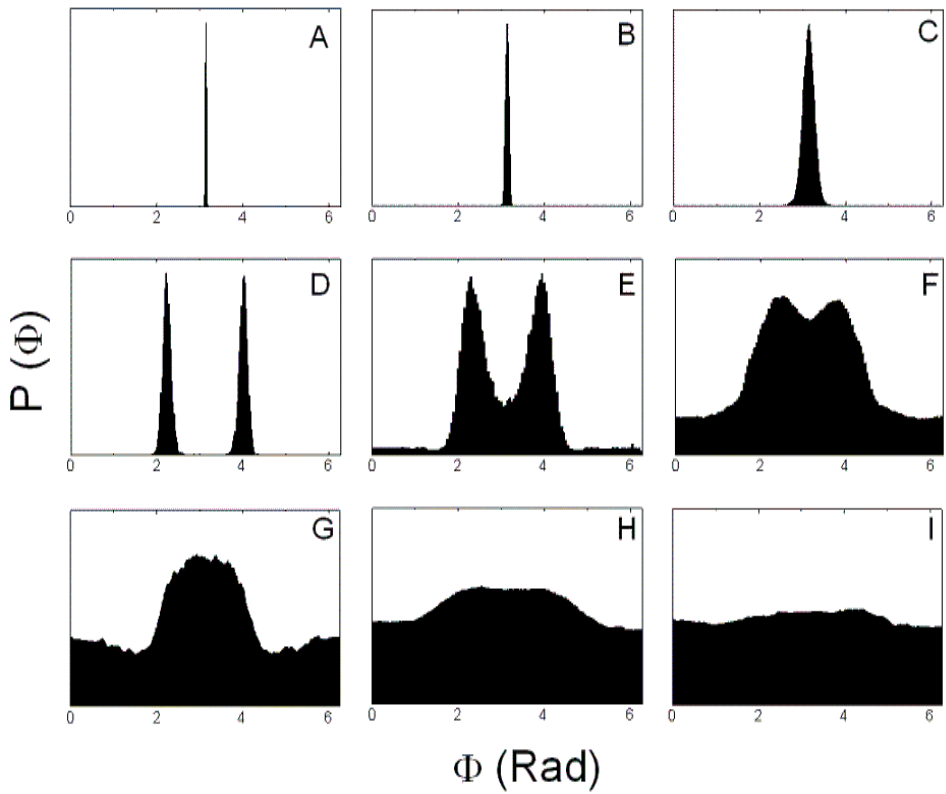}
\caption{The distribution of the cyclic relative phase $P(\Phi)$
corresponding to some representative points of the synchronization
diagram in Fig.~\ref{fig:5}. (A) $\varepsilon=0.3$, $N=2\times
10^{5}$; (B) $\varepsilon=0.3$, $N=2\times 10^{4}$; (C)
$\varepsilon=0.3$, N$=2\times 10^{3}$; (D) $\varepsilon=0.08$,
$N=2\times10^{5}$; (E) $\varepsilon=0.08$, $N=2\times 10^{4}$; (F)
$\varepsilon=0.08$, $N=2\times 10^{3}$; (G) $\varepsilon=0.02$,
$N=2\times 10^{4}$; (H) $\varepsilon=0.02$, $N=2\times 10^{3}$; (I)
$\varepsilon=0.004$, $N=2\times 10^{3}$. Each plane have different
vertical scales.} \label{fig:6}
\end{figure*}

Given a data set or some model dynamics there exists a variety of
methods to define an instantaneous phase $\phi(t)$ of a signal or a
dynamics \cite{Hanggi}. However, for a stochastic system it is
essential to assess the robustness of the phase definition with
respect to noise. In many practical applications like neural spike
sequence, it is useful to define an instantaneous phase $\phi(t)$ by
linear interpolation,
\begin{equation}
\phi(t)=2\pi\frac{t-t_{n}}{t_{n+1}-t_{n}}+2\pi n \quad
(t_{n}\leqslant t\leqslant{t_{n+1}}),
\label{eq-10}
\end{equation}
where $t_{n}$ is the time at which the neuron fires a spike. The
instantaneous phase difference between them is then given by
\begin{equation}
\psi(t)=\phi_{1}(t)-\phi_{2}(t).
\label{eq-11}
\end{equation}
Phase synchronization is a weak form of synchronization in which
there is a bounded phase difference of two signals. Usually, the
relative phase can vary from $-\infty$ to $+\infty$ in stochastic
system if the coupling is weak and/or the noise level is high.
However, if we increase coupling strength and adjust noise to a low
level, the relative phase will fluctuate around some constant
values. Sometimes, noise would induce a phase slip where the
relative phase changes abruptly by $\pm2\pi$. Thus, it is useful to
define the phase locking condition in a statistical sense by the
cyclic relative phase \cite{kurth}
\begin{equation}
\Phi=\psi(mod2\pi).
\label{eq-12}
\end{equation}
A dominant peak of the distribution of this cyclic relative phase
$P(\Phi)$ reflects the existence of a preferred relative phase for
the firing of both neurons. When this preferred phase is zero we
speak of phase synchronization in a statistic sense. We speak of
out-of-phase synchronization when the distribution $P(\Phi)$ peaks
around a nonzero value of $\Phi$. Especially if the nonzero value is
$\pi$, we call it antiphase synchronization \cite{Hanggi,Casado2}.

\subsection{phase synchronization of identical neurons\label{sec4-1}}

\begin{figure}
\includegraphics[width=0.5\textwidth]{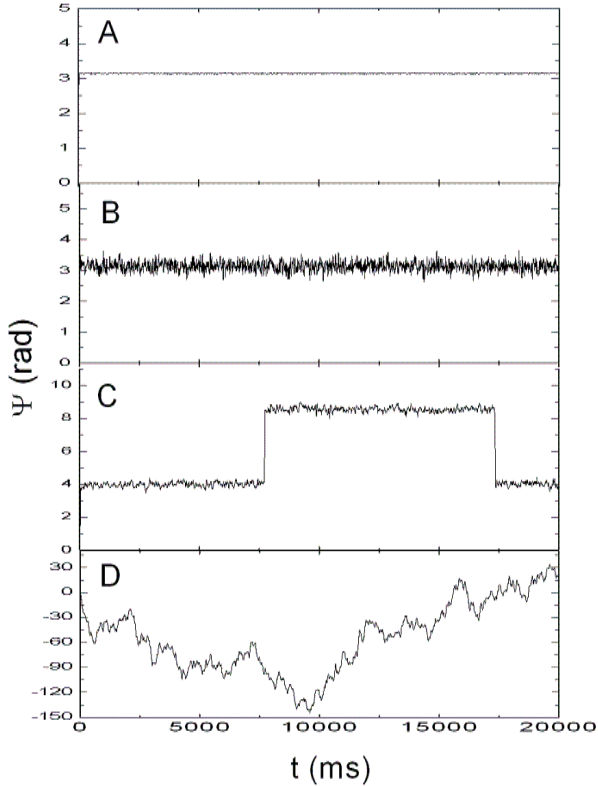}
\caption{Temporal evolution of the relative phase (A)
$\varepsilon=0.3$, $N=2\times 10^{5}$; (B) $\varepsilon=0.3$,
$N=2\times10^{3}$; (C) $\varepsilon=0.08$, $N= 10^{4}$; (D)
$\varepsilon=0.02$, $N=2\times 10^{3}$.} \label{fig:7}
\end{figure}

 In this section, we study phase synchronization of two identical neurons ($N_{1}=N_{2}=N$). In Fig.~\ref{fig:5}, we present the synchronization diagram in terms of the coupling strength $\varepsilon$ and channel cluster size $N$. A different form of the distribution $P(\Phi)$, which is plotted in Fig.~\ref{fig:6}, characterizes each region in it. The corresponding temporal evolution of the relative phase is illustrated in Fig.~\ref{fig:7}. We will give a detailed description of each region in the following part.

In region 1, the distribution shows a monomodal character as plotted
in panel A, B, C of Fig.~\ref{fig:6}. In this region, when both
$\varepsilon$ and $N$ are very large, the distribution of $P(\Phi)$
is a narrow peak on $\pi$. Thus we can speak of antiphase
synchronization for a statistic out-of-phase locking is achieved.
With the decreasing of $N$, the peak is still on $\pi$ but becomes
broader(see the change of peaks in $A\rightarrow B \rightarrow C$ ).
This suggests that in the case of large coupling strength, large
channel cluster size (i.e., small channel noise) allow the
statistical antiphase synchronization to approach full antiphase
synchronization appearing in deterministic systems. As can be seen
in Fig.~\ref{fig:5}, there is a minimal value of the coupling
strength $\varepsilon_{0}$ for which the antiphase locking becomes
stable in a statistical sense. This minimal coupling strength
$\varepsilon_{0}$ ($\approx$ 0.115) is independent of the channel
cluster size. To show the effect of channel noise on antiphase
synchronization, we demonstrated the temporal evolution of relative
phase in this situation in panel A and B of Fig.~\ref{fig:7}.
Obviously, decreasing channel cluster sizes will lead to larger
fluctuation of the relative phase due to channel noise, thus give
distribution of $P(\Phi)$ a broader peak, but it does not destruct
antiphase synchronization.

Region 2 marked by open circles corresponds to the bimodal
distribution of $P(\Phi)$, as shown in panel D, E and F of
Fig.~\ref{fig:6}. In this region, when $\varepsilon$  and $N$ are
large [2(a)], the two peaks of the distribution are well spaced from
each other.  To uncover the underlying mechanism of this bimodal
distribution, we investigated the temporal evolution of relative
phase in this situation. As we observed in panel C of
Fig.~\ref{fig:7}, $\psi(t)$ will fluctuates successively around one
of a pair of symmetric values for a long period then switch suddenly
to the other one. This fact clearly reflects the two-state character
of the phase dynamics. We argue that this two-state dynamics is the
result of a compromise between coupling and noise. The existence of
a two-state dynamics suggests the possibility of inducing a kind of
stochastic resonant behavior by coupling the relative phase to the
action of an external periodic forcing \cite{Casado2}. Again, if we
decrease $N$ to enter 2(b) area, the two peaks will be broader, and
gradually overlapped and have small wings. The overlapping means
that the bimodal distribution becomes unstable. The wings implies
that preferred relative phase for the firing of both neurons does
not prefer to some certain values anymore, thus the synchronization
becomes weak. If we further decrease $N$ to enter 2(c) area, we find
that the two peaks move closer and then merge but still have a
maxima around $\pi$ (see panel F of Fig.~\ref{fig:6}.)

\begin{figure}
\includegraphics[width=0.50\textwidth]{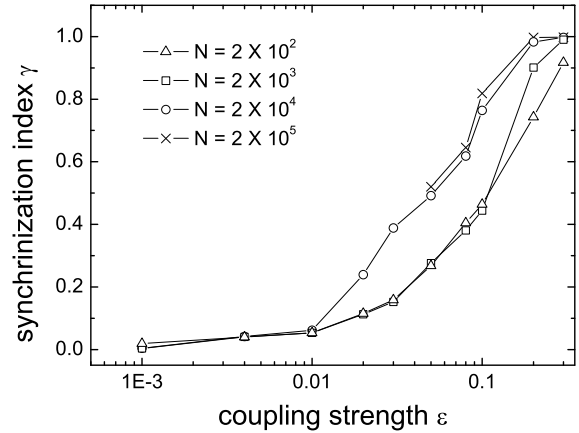}
\caption{Synchronization indices for two identical neurons \emph{versus} coupling strength  with various ion channel cluster sizes $N$.}
\label{fig:8}
\end{figure}

Actually, the phase-locking phenomena in noise-free neural systems
have been well studied through effective coupling analysis\cite{Seon
Hee Park 1, Seung,Seon Hee Park}. S. K. Han and Kuramoto had
demonstrated that diffusive interaction will dephase interacting
oscillators and may stabilize them at a phase difference given by
the corresponding stable fixed point according to the initial
condition (see detail in Ref.~\cite{Seung}). In our case, those
stable fixed points are stochastic variables with single peaks
distribution alike the peaks demonstrated in Fig.~\ref{fig:6}. In
region 1, there is only one stable fixed point distributed around
$\pi$ and will become broader if the noise intensity is increased.
In region 2, the system has two stable fixed points. The system will
be stabilized at one point according to the initial condition, then
the channel noise occasionally change the initial condition and
stabilized the system at the other one. If the channel cluster size
$N$ is extremely large, though there are still two stable fixed
point, the channel noise is too weak to change the initial condition
frequently, and only one of the two peaks can be observed with
certain recording time interval(not shown). As $N$ decreases, the
channel noise becomes larger, giving broader distributions of the
two stable fixed points and more frequent switches between them. It
is the broader distributions of the two stable fixed points that
leads the overlapping of the two peaks in this area.

In region 3, we find the bimodal distribution of $P(\Phi)$ mentioned
before disappears and the single peak distribution appears again
with large wings. Panel G of Fig.~\ref{fig:6} shows a representative
cyclic relative phase distribution corresponding to the 3(a) area,
which is characterized by a peak around $\pi$ and another smaller
one around $0$. If we decreasing the coupling strength to enter 3(b)
area from 2(c) area, the central peaks will decrease in height
[panel H ] and eventually disappear [3(c) area, panel I]. In 3(c)
area, the relative phase will drift unboundedly and $P(\Phi)$ ceases
to be useful (see panel D of Fig.~\ref{fig:7}). When the coupling
strengths is very weak, the system at hand can be considered as two
uncoupled neurons which fire independently due to ion channel noise.
Thus their relative phase can be at any arbitrary value (as show in
panel D of Fig.~\ref{fig:7}), and gives relative phase a smooth
distribution.

Region 4 in Fig.~\ref{fig:5} is the silent state in which both
neurons cannot fire spikes but only perform low amplitude
fluctuations around its resting potentials under the combining
effects of coupling and channel noise.

 Next, we characterize those peaks with
synchronization indices which are defined as
\begin{equation}
\gamma^{2}=\langle cos\Phi(t)\rangle^{2}+\langle sin\Phi(t)\rangle^{2},
\label{eq-9}
\end{equation}
where $\langle ...\rangle$ denotes temporal averaging. The index
$\gamma$ assumes values between 0 (no synchronization) and 1
(perfect phase locking)\cite{Hauschild}.

Fig.~\ref{fig:8} quantitatively demonstrated the synchronization of
two identical neurons under the effect of both coupling strength and
channel noise. When coupling strength is small, the two neurons show
almost no synchronization [$\gamma\approx0$, corresponding to 3(c)
area of Fig.~\ref{fig:5}] or silent state when $N$ is
large(incomplete  lines, corresponding to region 4 of
Fig.~\ref{fig:5}). The synchronization index $\gamma$ is not
sensitive to the change of channel cluster size $N$ in small
coupling strength region. As the coupling strength increases the
degree of synchrony of the two neurons increases. The maximal
synchrony appears at $\gamma\approx1$ (corresponding to region 1 of
Fig.~\ref{fig:5}). Note that when channel cluster sizes are large,
as shown in Fig.~\ref{fig:5}, there exists a step-like transition (a
threshold) to perfect synchronization. The threshold is around about
$\varepsilon=0.04$ when $N=2\times10^{5}$. However, it disappears
when cluster sizes decrease, and the transition becomes a graded
type. It implies that channel noise can 'soft' the threshold to give
a wider range of synchronization degree.

\subsection{phase synchronization of nonidentical neurons\label{sec4-2}}

\begin{figure}
\includegraphics[width=0.50\textwidth]{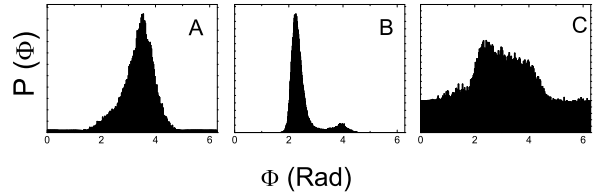}
\caption{The distribution of the cyclic relative phase $P(\Phi)$ for
nonidentical neurons. (A)$\varepsilon=0.2, N_{1}=2\times10^{2}$,
$N_{2}=2\times10^{3}$; (B)$\varepsilon=0.08, N_{1}=1\times10^{5}$,
$N_{2}=2\times10^{4}$; (C)$\varepsilon=0.03, N_{1}=2\times10^{2}$,
$N_{2}=2\times10^{3}$. Each plane have different vertical scales.}
\label{fig:9}
\end{figure}

\begin{figure}
\includegraphics[width=0.50\textwidth]{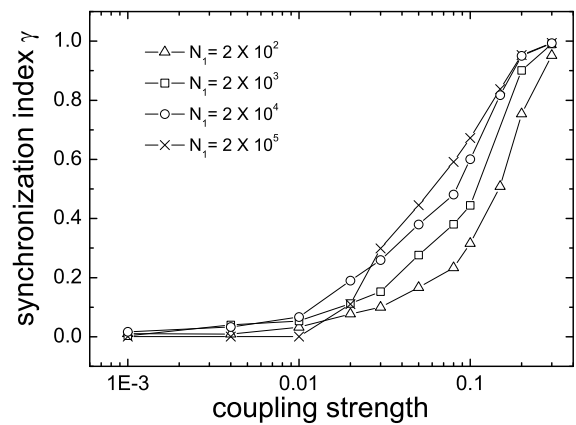}
\caption{Synchronization indices for two nonidentical neurons
\emph{versus} coupling strength with channel cluster size of the
first neuron $N_{2}=2\times10^{3}$.} \label{fig:10}
\end{figure}

Actually, neurons in nature are not identical. The nonidentity can be
 achieved in numerical simulation by mismatching neuronal parameters
  (the leakage conductance $g_{l}$, for example) \cite{Changsong Zhou}.
   Here, we introduce a mismatch into the channel cluster sizes of the
    two neurons (i.e., $N_{1} \neq N_{2}$). With this parameter heterogeneity,
     the neuron with smaller channel cluster size, due to its larger channel
     noise, is easier to be excited by subthreshold stimuli and has larger
      firing rate than another one.

In the case of two nonidentical neurons, the above mentioned cyclic
relative phase distributions are still tenable and the perfect phase
synchronization can also be achieved (see Fig.~\ref{fig:10}).
However, there are three exceptions. First, because the symmetry of
the distribution $P(\Phi)$ is dependent on the symmetry of the
system, for nonidentical neurons the distribution of the cyclic
relative phase is asymmetric (see Fig.~\ref{fig:9}). This fact was
also confirmed by applying different tonic subthreshold currents to
two neurons to make the system asymmetric \cite{Casado2}. Second, as
mentioned before, the two weakly coupled identical neurons with
large channel cluster sizes are unable to fire spikes under a
subthreshold stimulus (see the plot of $N=2\times 10^{5}$ in
Fig.~\ref{fig:8}). However, when a neuron with large channel cluster
size coupled with a small one which could be excited by subthreshold
stimulus due to channel noise, the large one is excited by the small
one through coupling. As shown in Fig.~\ref{fig:10}, comparing with
identical situation, the neuron can be excited when $N=2\times
10^{5}$ and $\varepsilon$ is rather small. It is also seen that in
the weak coupling region, the identical neurons exhibit higher
degree of synchronization than the nonidentical ones. Whereas in the
strong coupling region, when a neuron is coupled to another one
which has larger $N$, they exhibit higher degree of synchronization.
This is consistent with the frequency synchronization case where
identical neuron is frequency locked even coupling is weak, but
nonidentical neuron can also be synchronized if the coupling is
strong, and neurons with large channel clusters are easier to get
synchronized.

\section{CONCLUSION\label{sec5}}

In conclusion, the frequency and phase synchronization of two
coupled stochastic Hodgkin-Huxley neurons are studied by varying
coupling strength and channel cluster sizes. The two neuron is
coupled via a gap junction because the gap-junctional (diffusive)
coupling can generate rich dynamical behavior \cite{Sherman}. What's
more, with this simple coupling, we could emphasized on the effects
of channel noise and ignore the inessential details of complex
synaptic process. Our studies show that when the coupling is weak,
the cluster sizes of the two neurons must be the same to achieve
frequency synchronization, and the synchronization region is very
narrow. However, when coupling is strong, the two neurons can be
frequency entrained in a wide region. For two identical neurons, a
state of statistical antiphase synchronization is reached in the
strong coupling region if the cluster size is large enough. In this
state, the relative phase between two spike trains would be around
$\pi$. As the coupling strength and channel cluster size are
reduced, the phase-locking condition is lost and a rather complex
behavior would appear. This complex behavior, as we have argued, is
the result of a compromise between coupling strength and channel
noise. We use synchronization indices to characterize the transitions to synchronization,
and find that there exit a threshold to synchronization.
When channel cluster sizes are small, channel noise can 'soft' this
threshold to present synchronization at a wider range. For two
nonidentical neurons, the distribution of the cyclic relative phase
is asymmetric and the silent state in identical situation
disappears. This, as we have pointed out, is due to asymmetric of
the system and spontaneous firing induced by channel noise.

It is helpful that our study is important for the understanding of
coupled stochastic systems and possible applications especially in
neuroscience where the synchronization activity could be tuned
through the control of the channel noise via channel blocking. By
applying channel cluster size control to real neural systems one
should be able to influence neural synchrony. Further work should
focus on more sophisticated models and on coupling more than two
neurons.

\section*{Acknowledgments}
We really appreciate two anonymous referees for their very constructive and helpful suggestions. This work was supported by the National Natural Science Foundation of China with Grant No. $10305005$ and by the Special Fund for Doctor Programs at Lanzhou University.

\end{document}